\documentclass[12pt, amsfonts]{article}
\begin{document}
\def\p {{\partial}}
\def\n {{\nu}}
\def\m {{\mu}}
\def\a {{\alpha}}
\def\bt {{\beta}}
\def\f {{\phi}}
\def\th {{\theta}}
\def\g {{\gamma}}
\def\eps {{\epsilon}}
\def\e {{\psi}}
\def\k {{\chi}}
\def\la {{\lambda}}
\def\na {{\nabla}}
\def\bn {\begin{eqnarray}}
\def\en {\end{eqnarray}}
\title{Reduced phase-space quantization of constrained systems\footnote{e-mail:
$sami_{-}muslih$@hotmail.com}} \maketitle
\begin{center}
\author{S.I. MUSLIH\\ {\it Deptartment of Physics, Al-Azhar University,
Gaza, Palestine}}
\end{center}

\begin{abstract}
The Hamilton - Jacobi method of constrained systems is discussed.
The equations of motion for three singular systems are obtained
as total differential equations in many variables. The
integrability conditions for these systems lead us to obtain the
canonical reduced phase space coordinates with out using any gauge
fixing condition. The operator and the path integral quantization
of these systems are discussed.
\end{abstract}

\newpage

\section{Introduction}

Recently, the canonical method [1-4] has been developed to
investigate constrained systems. The equations of motion are
obtained as total differential equations in many variables which
require the investigation of integrability conditions. If the
system is integrable, one can solve the equations of motion
without using any gauge fixing conditions.

Now we would like to give a brief discussion of the canonical
method. This method gives the set of Hamilton - Jacobi partial
differential equations [HJPDE] as

\bn
&&H^{'}_{\a}(t_{\bt}, q_a, \frac{\p S}{\p q_a},\frac{\p S}{\p
t_a}) =0,\nonumber\\&&\a, \bt=0,n-r+1,...,n, a=1,...,n-r,\en where
\begin{equation}
H^{'}_{\a}=H_{\a}(t_{\bt}, q_a, p_a) + p_{\a},
\end{equation}
and $H_{0}$ is defined as
\bn
 &&H_{0}= p_{a}w_{a}+ p_{\m} \dot{q_{\m}}|_{p_{\n}=-H_{\n}}-
L(t, q_i, \dot{q_{\n}},
\dot{q_{a}}=w_a),\nonumber\\&&\m,~\n=n-r+1,...,n. \en

The equations of motion are obtained as total differential
equations in many variables as follows:

\bn
 &&dq_a=\frac{\p H^{'}_{\a}}{\p p_a}dt_{\a},\;
 dp_a= -\frac{\p H^{'}_{\a}}{\p q_a}dt_{\a},\;
dp_{\bt}= -\frac{\p H^{'}_{\a}}{\p t_{\bt}}dt_{\a}.\\
&& dz=(-H_{\a}+ p_a \frac{\p
H^{'}_{\a}}{\p p_a})dt_{\a};\\
&&\a, \bt=0,n-r+1,...,n, a=1,...,n-r\nonumber \en where
$z=S(t_{\a};q_a)$. The set of equations (4,5) is integrable [3,4]
if

\begin{equation}
dH^{'}_{0}=0,\;\;\; dH^{'}_{\m}=0,  \m=n-p+1,...,n.
\end{equation}
If condition (6) are not satisfied identically, one considers
them as new constraints and again testes the consistency
conditions. Hence, the canonical formulation leads to obtain the
set of canonical phase space coordinates $q_a$ and $p_a$ as
functions of $t_{\a}$, besides the canonical action integral is
obtained in terms of the canonical coordinates.The Hamiltonians
$H^{'}_{\a}$ are considered as the infinitesimal generators of
canonical transformations given by parameters $t_{\a}$
respectively.

\section{Quantization of constrained systems}

For the quantization of constrained systems we can use the
Dirac's method of quantization [5,6], or the path integral
quantization method [7,8].

Now will shall give a brief information about these two methods.
\subsection{Operator quantization}

For the Dirac's quantization method we have
\begin{equation}
H^{'}_{\a}\Psi=0,\;\;\;\a=0,n-r+1,...,n,
\end{equation}
where $\Psi$ is the wave function. The consistency conditions are
\begin{equation}
[H'_{\m}, H'_{\n}]\Psi=0,\;\;\;\m,\n=1,...,r,
\end{equation}
where$[,]$ is the commutator. The constraints $H'_{\a}$ are
called first- class constraints if they satisfy
\begin{equation}
[H'_{\m}, H'_{\n}]=C_{\m\n}^{\g}H'_{\g}.
\end{equation}

In the case when the Hamiltonians $H'_{\m}$ satisfy
\begin{equation}
[H'_{\m}, H'_{\n}]=C_{\m\n},
\end{equation}
with $C_{\m\n}$ do not depend on $q_{i}$ and $p_{i}$, then from
(8) there arise naturally Dirac' brackets and the canonical
quantization will be performed taking Dirac's brackets into
commutators.

\subsection{Path integral quantization method}

The path integral quantization is an alternative method to perform
the quantization of constrained systems.

Now we shall give a brief review of the canonical path integral
formulation of constrained systems [7,8].

If the set of equations (4) is integrable then one can solve them
to obtain the canonical phase-space coordinates as
\begin{equation}
q_{a}\equiv q_{a}(t, t_{\m}),\;\;\;p_{a}\equiv p_{a}(t,
t_{\m}),\;\;\m=1,...,r,
\end{equation}
In this case, the path integral representation may be written as
[7,8]

\bn &&\langle Out|S|In\rangle=\int \prod_{a=1}^{n-r}dq^{a}~dp^{a}
\exp [i \{\int_{t_{\a}}^{{t'}_{\a}}(-H_{\a}+ p_a\frac{\p
H^{'}_{\a}}{\p p_a})dt_{\a}\}],\nonumber\\&&a=1,...,n-r,
\;\;\;\a=0,n-r+1,...,n. \en

One should notice that the integral (12) is an integration over
the canonical phase - space coordinates $(q_a, p_a)$.

\section{Examples}

As a first example we shall treat the relativistic particle as a
constrained system and demonstrate the fact that the gauge fixing
problem is solved naturally if the canonical path integral method
is used.

Let us consider the action of the a relativistic particle as

\begin{equation}
 S =\frac{1}{2}\int (\frac{{\dot x^{\m}}{\dot
x_{\m}}}{e} - e m^{2})d\tau,
\end{equation}

where $x^{\m}(\tau)$ and $e(\tau)$ are even variables. The
canonical momenta are defined as

\begin{equation}
p_{\m}= \frac{\p L}{\p {\dot x^{\m}}}=\frac{1}{e}(\dot
x_{\m}),\;\;\pi_{e}= \frac{\p L}{\p {\dot e}}= 0= -H_{1}.
\end{equation}

The canonical Hamiltonian $H_{0}$ can be obtained as

\begin{equation}
H_{0}=p_{\m}{\dot x^{\m}}- {\dot e}H_{1} - L=+\frac{e}{2}(p^2 +
m^2).
\end{equation}

Making use of equations (1,2) and (14, 15), the set of Hamilton-
Jacobi partial differential equations reads

\bn H'_{0}=&& p^{(\tau)} + H_{0}=0;\;\; p^{(\tau)}=\frac{\p
S}{\p \tau},\\
H'_{1}=&& \pi_{e}= 0;\;;\;\;\;\;\; \pi_{e}=\frac{\p S}{\p e},\\
\en
This set leads to the total differential equations as

\begin{equation}
 dx_{\m}=(e p_{\m})d\tau,\;dp_{\m}=0,\;d\pi_{e}= -\frac{1}{2}(p^2 + m^2)d\tau=0,
 \;dp^{(\tau)}=0.
\end{equation}

To check whether the set of equations (19) is integrable or not
let us consider the total variations of $ H'_{0}$ and $H'_{1}$. In
fact
\begin{equation}
dH'_{1} = -\frac{1}{2}(p^2 + m^2)d\tau=0 =H'_{2}d\tau.
\end{equation}
The total differentials of ${H'}_{0}$ and ${H'}_{2}$ vanish
identically, the equations of motion are integrable and the
canonical phase space coordinates $(q_{\m}, p_{\m})$ are obtained
in terms of parameters $(\tau, e).$

To obtain the operator quantization of this system one can follow
the procedure discussed in section (2.1). In this case one takes
the constraint equation as an operator whose action on the allowed
Hilbert space vectors is constrained to $zero$, i., e., $
H'_{2}\Psi =0$, we obtain
\begin{equation}
[{\hat p}^{\m}{\hat p}_{\m} + m^{2}]\Psi=0,
\end{equation}

Now to obtain the path integral quantization of this system, we
can use equation (5) to obtain the canonical action as
\begin{equation}
S=\int \frac{e}{2}(p^2 - m^2)d\tau.
\end{equation}

Making use of (22) and (12) the path integral for the system (13)
is obtained as
\begin{equation}
\langle q_{\m},e, \tau; {q'}_{\m},e', {\tau}^{'}\rangle=
\int_{q_{\m}}^{{q'}_{\m}}\prod_{\m} dq^{\m}~dp^{\m}\exp
[i\{\int_{{\tau}}^{{\tau}'} \frac{e}{2}(p^2 - m^2)d\tau\}].
\end{equation}
This path integral representation is an integration over the
canonical phase space coordinates $q_{\m}$ and $p^{\m}$.

Now for a system with $ n$ degrees of freedom  and $ r$ first
class constraints $\f^{\a}$, the matrix element of the $S$ -
matrix is given by Faddeev and Popov [9,10] as
\begin{equation}
\langle Out|S|In\rangle=\int \prod_{t} d{\m}(q_j, p_j)\exp [i
\{\int_{-\infty}^{\infty}dt(p_j \dot{q_j} - H_0)\}],\;\;j=1,...,n,
\end{equation}
where the measure of integration is given as
\begin{equation}
d {\m}= det|\{\f^{\a},
\k^{\bt}\}|\prod_{\a=1}^{r}\delta(\k^{\a})\delta(\f^{\a})\prod_{j=1}^{n}dq^{j}
dp_{j},
\end{equation}
and $\k_{a}$ are $r$- gauge constraints.

If we perform now the usual path integral quantization [9,10]
using (24) for system (13), one must choose two  gauge fixing
conditions to obtain the path integral quantization over the
canonical phase-space coordinates.

As a second example, we consider the Lagrangian
\begin{equation}
L(q_{1}, q_{2}, {\dot{q_{1}}}, {\dot{q_{2}}}) =
\frac{{\dot{q_{1}}}^{2}}{4q_{2}} - q_{2}({q_{1}}^{2} +
\frac{{q_{2}}^{2}}{3} - R^{2}),
\end{equation}
excluding the line $q_{2}=0$ on the configuration space. The
canonically conjugated momenta are obtained as \bn&&p_{1}=
\frac{\dot{q_{1}}}{2q_{2}},\\
&&p_{2}=0. \en The canonical Hamiltonian can be obtained as
\begin{equation}
H_{0}= q_{2} {p_{1}}^{2} + q_{2}({q_{1}}^{2}
+\frac{{q_{2}}^{2}}{3} - R^{2}).
\end{equation}
Making use of eqns. (1,2) and (28,29), the set of HJPDE reads

\bn&& {H'}_{0}= p_{0} + H_{0}=0,\\
&&{H'}_{1}= p_{2}=0, \en where $ p_{0}= \frac{\p S}{\p t}$,
$p_{2}=\frac{\p S}{\p q_{2}}$, here $S = S(q_{1}, q_{2}, t)$
represents the action.

This set leads to the following total differential equations

\bn&&dq_{1}= 2p_{1}q_{2} dt,\\
&&dp_{1}= -2q_{1}q_{2} dt,\\
&&dp_{2} = -({q_{1}}^{2} + {q_{2}}^{2} + {p_{1}}^{2} - R^{2})dt=0.
\en According to eqns. (6) and (31,34) the vanishing of the total
differential of ${H'}_{1}$ leads to the constraint
\begin{equation}
{H'}_{2}={q_{1}}^{2} + {q_{2}}^{2} + {p_{1}}^{2} - R^{2}.
\end{equation}
Since ${H'}_{2}$ is not identically zero, we consider it as new
constraint. Thus for a valid theory, total variation of ${H'}_{2}$
should be zero. Thus one gets
\begin{equation}
dq_{2}=0,
\end{equation}
which has the following solution
\begin{equation}
q_{2}= c= \pm{\sqrt{R^{2} -{p_{1}}^{2}- {q_{1}}^{2}}},
\end{equation}
where $c$ is an arbitrary constant.

Now the set of equations (32-34) is integrable and the canonical
phase space coordinates $(q_{1}, p_{1})$ are obtained as
functions of $(t, q_{2}= c)$ and in this case it is disc of
radius $R$ without the boundary. Choosing $q_{2}$ to be positive,
the canonical Hamiltonian can be written as
\begin{equation}
H_{0}= \frac{2}{3}( R^{2} - {p_{1}}^{2} - {q_{1}}^{2})^{3/2}.
\end{equation}

To proceed the canonical quantization of this system one can
follow the procedure discussed in section (2.1) . In this case one
takes the constraint equation as an operator whose action on the
allowed Hilbert space vectors is constrained to $zero$, i., e., $
H'_{0}\Psi(q_{1}) =0$, we obtain
\begin{equation}
[{\hat p}_{0} +  \frac{2}{3}( R^{2} - {\hat p_{1}}^{2} - {\hat
q_{1}}^{2})^{3/2}]\Psi(q_{1})=0,.
\end{equation}
in his case, the Hilbert space ${\cal H}$ consists of square
integrable functions in the interval $-R < q_{1}< R$.

Now to obtain the path integral quantization of this system, we
can use equation (5) to obtain the canonical action as
\begin{equation}
S=\int [2{p_{1}}^{2}{\sqrt{R^{2} -{p_{1}}^{2}-
{q_{1}}^{2}}}-\frac{2}{3}( R^{2} - {p_{1}}^{2} -
{q_{1}}^{2})^{3/2}]dt .
\end{equation}

Making use of (40) and (12) the path integral for the system (26)
is obtained as \bn\langle q_{1},t ; {q'}_{1},t'\rangle=&&
\int_{q_{1}}^{{q'}_{1}}\prod dq_{1}~dp_{1}\exp [i\{\int_{t}^{{t'}}
(2{p_{1}}^{2}{\sqrt{R^{2} -{p_{1}}^{2}- {q_{1}}^{2}}}\nonumber\\
&&-\frac{2}{3}( R^{2} - {p_{1}}^{2} - {q_{1}}^{2})^{3/2})dt\}].
\en

Consider next, the third Lagrangian
\begin{equation}
L(q_{1}, q_{2}, {\dot{q_{1}}}, {\dot{q_{2}}}) =
\frac{{\dot{q_{1}}}^{2}}{4q_{2}} - q_{2}({q_{1}}^{2} -
\frac{{q_{2}}^{2}}{3} - R^{2}),
\end{equation}
The canonical Hamiltonian can be obtained as
\begin{equation}
H_{0}= q_{2} {p_{1}}^{2} + q_{2}({q_{1}}^{2}
-\frac{{q_{2}}^{2}}{3} - R^{2}).
\end{equation}
The set of HJPDE reads

\bn&& {H'}_{0}= p_{0} + H_{0}=0,\\
&&{H'}_{1}= p_{2}=0. \en  As for previous system, the equations
of motion are integrable and the canonical phase space
coordinates $(q_{1}, p_{1})$ are obtained as functions of $(t,
q_{2}= c)$. Here $q_{2}$ has to branches as
\begin{equation}
q_{2}= c= \pm{\sqrt{{p_{1}}^{2}+ {q_{1}}^{2} -R^{2}}},
\end{equation}
where $c$ is an arbitrary constant. For each choice the canonical
phase space is $2$-dimensional infinite plane with a hole of
radius $R$ at the center, where we restricted $q_{2}$ to be
positive.

This system can be quantized using the method of section (2.1).
The Hilbert space consists of square-integrable functions on real
line ${\cal R}^{1}$ excluding the interval $[-R, R]$. Hence, we
obtain
\begin{equation}
[{\hat p}_{0} +  \frac{2}{3}( {\hat p_{1}}^{2} + {\hat
q_{1}}^{2}-R^{2})^{3/2}]\Psi(q_{1})=0.
\end{equation}

The the path integral for the system (42) is obtained as
\bn\langle q_{1},t ; {q'}_{1},t'\rangle=&&
\int_{q_{1}}^{{q'}_{1}}\prod dq_{1}~dp_{1}\exp [i\{\int_{t}^{t'}
(2{p_{1}}^{2}{\sqrt{{p_{1}}^{2}+ {q_{1}}^{2}-R^{2}}}\nonumber\\
&&-\frac{2}{3}( {p_{1}}^{2} + {q_{1}}^{2}-R^{2})^{3/2})dt\}]. \en
One should notice that this path integral is an inegration over
the canonical phase space coordinates $q_{1}$ and $p_{1}$.

\section{ Conclusion}

In this work we have obtained the quantization for three singular
systems. In the relativistic particle problem, (13). The
integrability conditions $ dH'_{0}$ and $dH'_{1}$ are satisfied,
the system is integrable. Hence the canonical phase space
coordinates $q_{\m}$ and $p_{\m}$ are obtained in terms of
parameters $\tau$ and $e$. Although $e$ is introduced as a
coordinate in the Lagrangian the presence of the constraints and
the integrability conditions forces as to treat it as a parameter
like $\tau$. In this case the Hamiltonians
 $H'_{0}$ and $H'_{1}$ are considered as infinitesimal
 generators of canonical transformations given by parameters
 $\tau$ and $e$ respectively and the path integral is obtained
 directly as an integration over the canonical phase space coordinates
 $q_{\m}$ and $p_{\m}$ without using any gauge fixing conditions.

 When applying the  Faddeev and Popov method to this
 model one has to choose two gauge fixing of the form and to integrate over the extended phase space coordinates
 and after integration over the redundant variables one can arrive at the result (23).

For the second and the third Lagrangians, these systems are
integrable and leads to obtain $dq_{2}= 0.$ Hence, the canonical
phase space coordinates $(q_{1}, p_{1})$ are obtained as
functions of $(t, q_{2}=constant)$. In this case the path
integral follows directly as an integration over the canonical
phase-space coordinates $(q_{1}, p_{1})$. In the usual
formulation [11], one has to integrate over the extended phase
space coordinates $(q_{1}, p_{1}, q_{2}, p_{2})$ and one can get
red of redundant variables $(q_{2}, p_{2})$ by using delta
functions $( \delta(p_{2}), \delta({q_{2}}^{2}+ {q_{1}}^{2}+
{p_{1}}^{2} -R^{2}))$ for the second system and the delta
functions $( \delta(p_{2}), \delta({q_{2}}^{2}+ R^{2}
-{q_{1}}^{2}- {p_{1}}^{2} ))$ for the third system.

 Unlike conventional methods one can  perform the path integral quantization
 of this system using
 the canonical path integral method to obtain the action directly without
 considering any Lagrange multipliers and without using delta functions in
 the measure.

\end{document}